# Automated discovery of GPCR bioactive ligands

**Author:** Sebastian Raschka (sraschka@wisc.edu)

**Address:** Department of Statistics, University of Wisconsin-Madison; 1300 Medical Sciences Center, Madison, Wisconsin 53706, USA

## Abstract

While G-protein coupled receptors (GPCRs) constitute the largest class of membrane proteins, structures and endogenous ligands of a large portion of GPCRs remain unknown. Due to the involvement of GPCRs in various signaling pathways and physiological roles, the identification of endogenous ligands as well as designing novel drugs is of high interest to the research and medical communities. Along with highlighting the recent advances in structure-based ligand discovery, including docking and molecular dynamics, this article focuses on the latest advances for automating the discovery of bioactive ligands using machine learning. Machine learning is centered around the development and applications of algorithms that can learn from data automatically. Such an approach offers immense opportunities for bioactivity prediction as well as quantitative structure-activity relationship studies. This review describes the most recent and successful applications of machine learning for bioactive ligand discovery, concluding with an outlook on deep learning methods that are capable of automatically extracting salient information from structural data as a promising future direction for rapid and efficient bioactive ligand discovery.

Introduction

G protein-coupled receptors (GPCRs), an important family of integral membrane proteins, play a crucial role in cellular signaling pathways in eukaryotes, and are among the most studied drug targets: GPCRs are the targets of approximately 34% of all drugs approved by the US Food and Drug Administration [1]. However, based on current estimates, only about 10% of known GPCRs are considered to be targeted by drugs for the treatment of a variety of human diseases, including hypertension, glaucoma, schizophrenia, and depression [2]. While the exact size of the GPCR family is unknown, genome analyses suggest that the GPCR family comprises approximately 800-1000 genes in humans [3,4]. More than 150 remain orphan receptors [5], meaning GPCRs for which endogenous ligands are still unknown. Since GPCRs are involved in many different cellular and biological processes and make excellent drug targets, the prediction and consequent identification of GPCR bioactive ligands is a topic of high interest and active research.

GPCR ligands differ in shape, size, and physicochemical properties and include proteins, peptides, lipids, steroids, and other small organic molecules [6]. Furthermore, GPCR ligands modulate receptor function in complex ways. Ligands exhibit a range of efficacy and can be categorized as full agonists, partial agonists, antagonists, or inverse agonists. GPCR ligand chemical diversity and functional complexity pose a challenge for discovering novel bioactive ligands and require experimental validation beyond what conventional and affordable binding affinity assays can offer. Thus, as a cost-effective alternative to wet-lab techniques such as high throughput screening (HTS) for identifying putative binding partners for more elaborate bioactivity assays, computational methods for ligand discovery have increased in popularity. Aside from being cost-intensive, an often-considered downside of HTS is the limited size and diversity of available ligand libraries compared to freely available computational libraries, which contain up to hundreds of millions of molecules [7–9].

Virtual screening has become a major approach for computer-aided ligand discovery and is traditionally categorized as either ligand- or structure-based virtual screening. Ligand-based virtual screening does not require knowledge of the target structure (i.e., the receptor) and can be summarized as a similarity search to known ligands, based on the hypothesis that molecules similar to a known binder are also likely to bind the target receptor. Structure-based methods usually involve docking a ligand into a receptor's binding pocket and use a scoring function to rank a library of ligands by their predicted affinities. Assessment criteria besides similarity to a known target and docking scores of potential ligands include chemical diversity, interaction with key residues, and other more general chemical characteristics such as drug likeness [10], not matching any known pan-assay interference compounds (PAINS) [11], and exhibiting favorable absorption, distribution, metabolism, excretion, and toxicity (ADMET) profiles [12].

The main limitation of molecular docking, which particularly applies to GPCR targets, is the limited availability of target structures. While significant advancements in membrane protein X-ray crystallography and cryo-electron microscopy have been made in recent years [13,14], available GPCR structures only span four out of the six different GPCR classes (A, B, C, F), where structures of class A GPCRs (rhodopsin-like receptors) form the largest proportion [15]. Crystal structures of 44 unique GPCRs are now available, and 205 GPCR structures have been obtained in their ligand-bound (often inactive, inhibitor-bound) state [1]. While the structures of the seven transmembrane helices are primarily conserved across the six GPCR classes, GPCRs can differ substantially in helical deformation and across the intracellular and extracellular loop domains, with the latter, in most GPCRs, forming the orthosteric ligand binding site or providing access to ligand binding sites that are buried within the transmembrane bundle [16]. The sequence and structural diversity of the extracellular loops is related to the diversity of ligand binding sites, which poses challenges for compensating the lack of experimental target structures by homology modeling in structure-based ligand discovery approaches. In the absence of structural information, non-structure-based approaches constitute the only viable alternative [17].

Machine learning, a field centered around the development and applications of algorithms that can learn from data automatically [18], is applicable in both ligand- and structure-based virtual screening. The subcategory of *supervised* machine learning (referred to as machine learning for the remainder of this article) focuses on algorithms that learn predictive models from examples. Hence, machine learning becomes particularly attractive as activity data becomes available after initial rounds of virtual screening and experimental assays, to guide further rounds of screening and experimental testing [19]. For instance, machine learning models can be used to predict the activity of untested molecules against a target receptor based on the learned relationship between tested molecules and their assay values (Figure 1), such as binding affinities, potencies, or binary labels (active/inactive) based on a practical threshold.

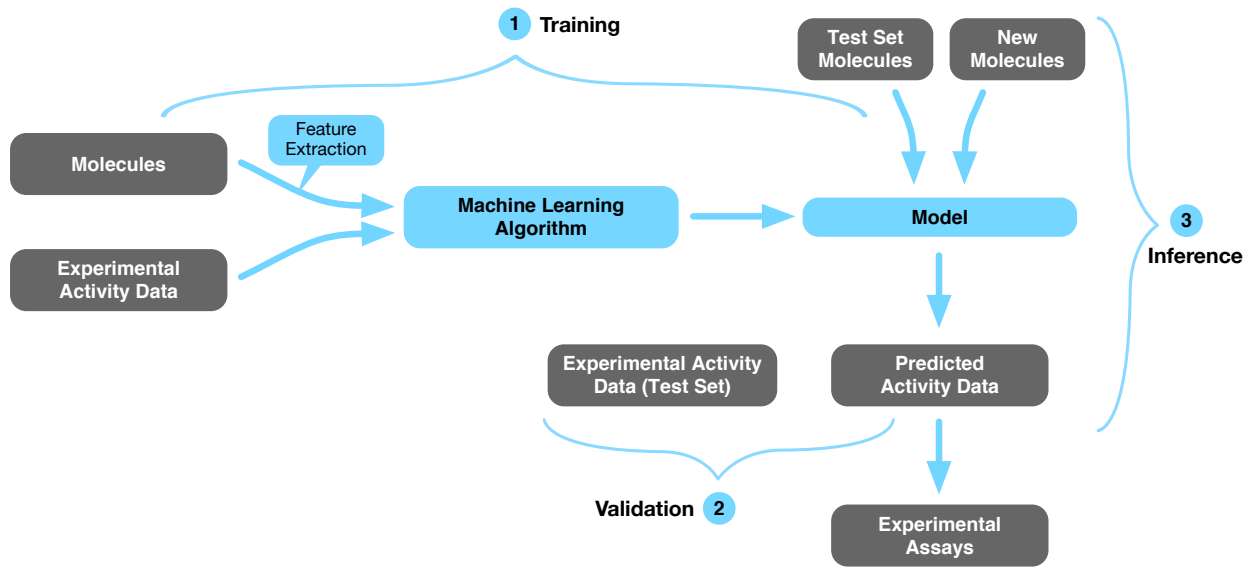

**Figure 1.** Illustration of a general, supervised machine learning workflow for the automated prediction of bioactive ligands of a GPCR. A machine learning algorithm learns a predictive model from examples of the inputs (representation of molecules) and outputs (activity data, which can be a continuous or categorical variable). Once the model was learned from training examples and (molecules associated with activity values) and properly evaluated, the model can be used to predict the activity of new molecules. The predicted activity can then be used to prioritize new molecules for experimental assays.

## Ligand-based virtual screening

As the structure determination of GPCRs is notoriously challenging [20], computational ligand discovery has traditionally focused on ligand-based approaches [21–23], which do not require knowledge of the structure of the target. Without knowledge about the molecular mechanism of interaction, though, molecules identified by ligand-based screening can exhibit agonistic or antagonistic properties, which must be determined experimentally. For instance, subtle differences in otherwise similar molecules identified by ligand-based virtual screening, such as changing a single keto-group of a full agonist into a hydroxyl group, can completely diminish the GPCR-mediated signaling response [17].

One of the biggest challenges of traditional ligand-based screening is the definition of a meaningful similarity measure [24–26] and the molecular representation, for example, one-dimensional, two-dimensional [27,28], or three-dimensional fingerprints representing physicochemical and/or structural features of the molecules [29] or representation of their complete 2D or 3D structures [30]. Even advanced similarity measures that consider volumetric as well as chemical similarity based on three-dimensional molecular overlays can be uncorrelated with the measured bioactivity of putative GPCR ligands [17]. Hence, in the absence of qualitative or quantitative structure-activity relationship (QSAR) models, similarity search-based approaches can be severely limited. For instance, in a recent benchmark study on 25 bioactivity datasets, researchers found that ligand similarity searching often does not perform better than random selection [31]. However, the same study showed that when a similarity search is combined with QSAR models, for example, using machine learning methods, the discovery rate of bioactive ligands can increase substantially.

### Structure-based virtual screening

Structure-based virtual screening for GPCR ligands has become more feasible in recent years due to increased structure quality and availability [32]. Since computational capabilities are rapidly advancing as well, docking studies are now often accompanied by molecular dynamics simulations, which allow more detailed studies of the GPCR-recognition process but remains infeasible large ligand libraries [33–36]. However, while the availability of experimental GPCR structures is growing, the currently available structures are mostly limited to inactive states. As of now, only a handful of atomic resolution GPCR structures are available in their fully active state [20]. This poses a challenge for structure-based approaches, as they suffer from a substantially lower performance when docking poses are sampled in the presence of non-ligand bound receptor structures [37].

The combination of molecular dynamics and docking can help identify key GPCR-ligand interactions on atomistic level [38] for designing pharmacophore models that can be utilized in structure-based, ligand-based, and machine learning-based identification of novel bioactive ligands. However, it should also be noted that while docking offers the advantage of gaining insights into ligand-receptor interactions for further study and ligand design, the ranking of binders according to affinity is often inaccurate. These methods are usually only capable of distinguishing between binders and non-binders [39]. Even in cases where docking leads to correct ranking among GPCR ligands according to their activity, the predicted patterns of interaction are often substantially different from the interactions found in crystal structures [40] and can be misleading for the further discovery and design of highly active ligands.

### Automated bioactive molecule discovery using machine learning

In recent years, machine learning has become one of the most widely used approaches in drug discovery and development [31,41–47]. Often, machine learning is combined with structure-based, ligand-based, and high-throughput screening to automate QSAR-based target prioritization in iterative and automated or semi-automated virtual screening pipelines (Figure 2).

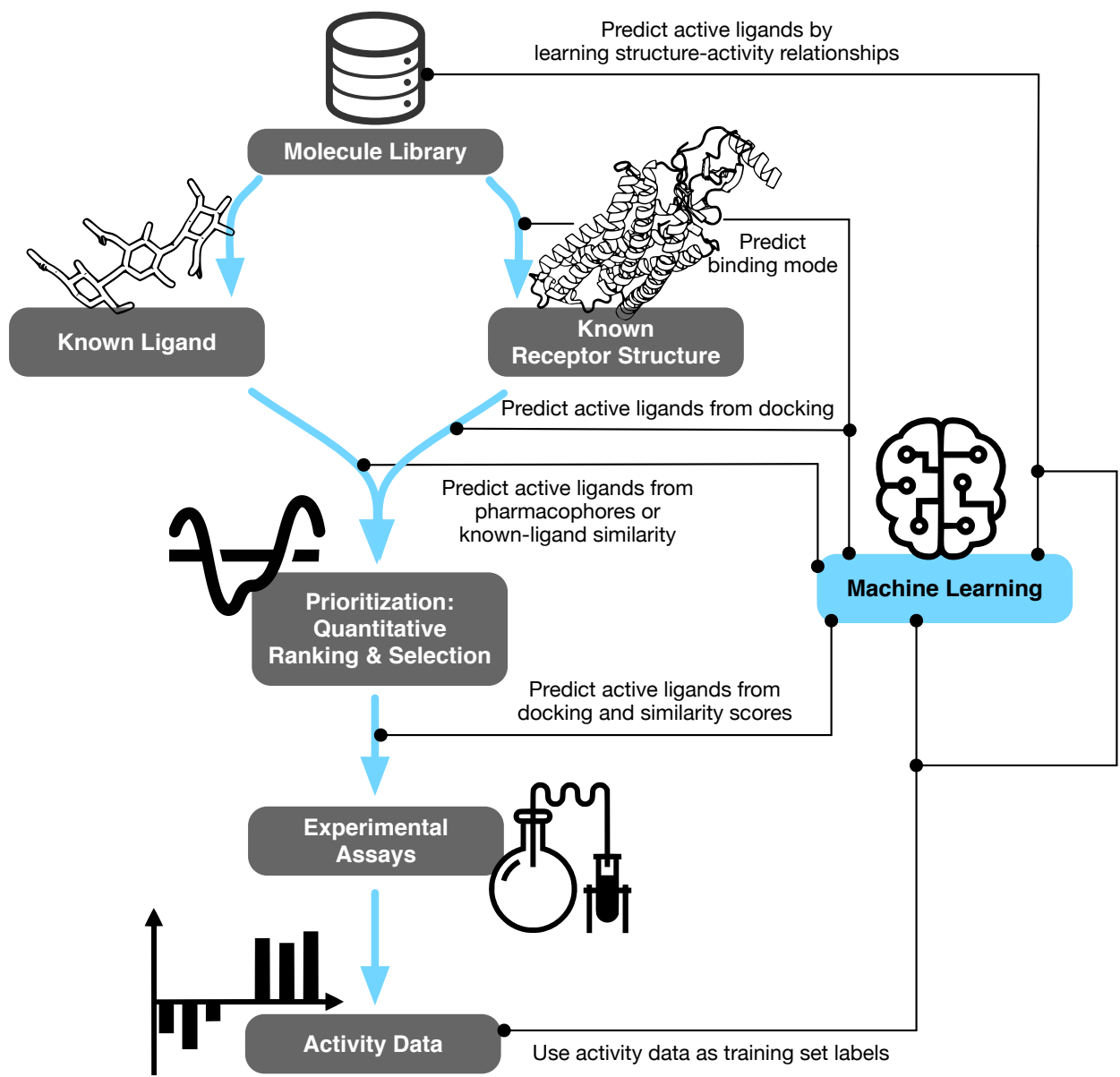

**Figure 2.** Illustration of how and where machine learning could be used to augment the discovery of bioactive ligands in a virtual screening pipeline. The gray boxes represent a virtual screening workflow that can be either ligand-based (if a bioactive ligand of the target receptor is known), structure-based (if the structure of the receptor, including the ligand binding side is known), or a combination thereof. The outputs (usually similarity or docking scores) are then used, together with domain expertise, to prioritize a set of ligands for experimental assaying from which activity data is obtained. Once activity data is available, machine learning algorithms can use these as training examples together with one or multiple forms of molecular representations to predict the bioactivity of new molecules that have not been tested against the target receptor, yet.

In a benchmark study including two GPCR targets, Ericksen *et al.* found that machine learning models trained on an ensemble of docking scores from multiple programs can boost the discovery rate of active ligands compared to ranking molecules from a single docking program [48]. Also, in the absence of the target GPCR structure, machine learning algorithms have been successfully used to discover active ligands based on functional group overlays with a known active from ligand-based virtual screening [19]. Furthermore, machine learning can be used to accurately predict target-ligand interactions, which was traditionally a task requiring docking and molecular dynamics simulations [49].

Moreover, machine learning offers the opportunity to predict the bioactivity of molecules where alternative methods such as docking are inappropriate due to the lack of high-quality experimental structures or homology models. Even in the absence of highly active molecules in training datasets, machine learning models can accelerate the discovery of highly active compounds, as a recent benchmark study based on 25 bioactivity datasets (including four GPCR targets) has shown [31].

While machine learning can automatically discover complex relationships in high-dimensional datasets that escape human interpretation, a common misconception is that machine learning-based models are uninterpretable black box models. Several methods exist to connect the predictions of a model with chemical features that explain bioactivity. These include model-specific approaches such as feature selection based on the weight coefficients of generalized linear models or calculating feature importance values in random forests based on information gain maximization as well as model-agnostic methods such as feature permutation evaluation [50], and LIME [51], among others.

For instance, applying feature selection and random forest algorithms to a dataset of molecular overlays of compounds, tested in assays to modulate the biological response of a GPCR involved in a pheromone signaling pathway, revealed that the presence of a sulfate group is a key requirement of bioactive molecules [19]. As described in their protocol [19], the researchers leveraged the fact that machine learning algorithms automatically learn complex relationships between molecular properties and experimentally measured bioactivity that maximize the prediction accuracy. The degree to which the model relied on the location of functional groups in inhibitor candidates (in relation to a known active) to drive the predictions was then used for the automated inference of bioactivity. Consequently, the researchers translated the automatically inferred functional group importance to filtering criteria for successive rounds of virtual screening, which lead to the discovery of additional actives as described in a related manuscript [17].

Machine learning, in particular the application of feature selection algorithms [50] combined with a nearest-neighbor classifiers, has also provided strong insights into which segments of GPCRs are flexible, independently rigid, or mutually rigid with other regions in active versus inactive GPCRs. This analysis by our group was recently carried out on the extracellular and intracellular loops and the N-terminal, central, and C-terminal segments of each of the helices in a series of different inactive and active GPCR structures. Assessing the flexible versus rigid state in all these segments by ProFlex [52] (htttps://github.com/psa-lab/ProFlex) followed feature importance analysis showed that the flexible versus rigid state of six segments alone could predict with high accuracy whether the GPCR was in an active or inactive state (96% accuracy for leave-one-out prediction across 27 GPCRs; Bemister-Buffington, Wolf, Raschka, and Kuhn, manuscript in preparation). We anticipate that this new method will also be useful for predicting whether designed ligands bound to GPCRs will behave as agonist or antagonists, based on the flexibility profile they induce in GPCRs.

Likely owed to the impressive results and state-of-the-art performance on complex tasks such as image analysis and language modeling, deep learning, a subfield of machine learning that focuses on the training of deep artificial neural networks, has emerged as the most recent trend and promising new direction for various molecular modeling tasks including ligand activity prediction and drug discovery [41,46,53]. When independent research groups compared different machine learning methods on various bioactivity datasets (including two GPCRs, dopamine D4 receptor and cannabinoid CB1 receptor), the results indicated that deep neural networks, while requiring more extensive tuning [54,55], generally outperform traditional off-the-shelf machine learning algorithms such as naïve Bayes classifiers [56–60], logistic regression [19,61–64], support vector machines [65,66], and random forests [67–70] when chemical fingerprints were used as input representations [54,55,71,72]. Also, one study found that deep neural networks outperform traditional machine learning methods on several ChEMBL bioactivity datasets across different molecular descriptors [73].

Although deep learning is the latest trend in machine learning and biological applications, the capacity and overparameterization of deep learning models have a higher tendency to overfit the training data and also require more extensive optimization. While machine learning algorithms generally benefit from larger dataset sizes [73], deep neural networks can be prone to overfitting in scenarios of small training sets and high correlation among the input features [53]. Focusing on estrogen receptor binding prediction, researchers found that traditional machine learning methods such as random forest and naïve Bayes are sufficient for predicting bioactive ligands [74,75], which may be related to the fact that deep neural network models tend to perform better on larger training sets and more complex chemical or physical feature representations [62,76].

As deep learning can be considered as a form of representation learning, the next logical step is to remove the need for feature engineering and fully automatically derive molecule descriptors that are best suited for a given target. For instance, Duvenaud *et al.* have demonstrated that neural networks can automatically learn feature representations that improve upon traditional manual fingerprint representations in various molecular prediction tasks [76]. While their study still relied on SMILES string inputs (a molecular description that encodes bond connectivity between atoms) to learn the molecular representation, graph convolutional neural network were recently used to learn representations directly from molecular graphs (where atoms are nodes and bonds are edges) for bioactivity prediction [73]. The approach outperformed standard fingerprints and representations constructed by chemoinformatics experts on various molecular property prediction tasks [77].

## Conclusions and perspectives

Even though machine learning, and especially the subfield of deep learning, raised skepticism in earlier years, companies and academics have now begun to embrace machine learning to further advance the automated discovery of drugs and other bioactive ligands [78]. This is partly also owed to the availability of free, open-source biological data science [79–81], machine learning, and deep learning software libraries [50,82–84], which make these technologies accessible to a wide audience. However, while automated methods for inference of bioactive ligands enable these discoveries, care should be taken that models make scientific sense and do not unintentionally exploit experimental artifacts [85]. Also, computational inferences do not replace the need for experimental assays for the validation and the study of biological effects; they simply enhance the selection of successful molecules. When molecules are discovered and prioritized for testing it is also essential to include extensive negative controls. For instance, we discovered that hydroxyl groups tend to lead to artificially high affinity scores in docking studies [86], resulting in false positives in ligand discovery.

While the recent progress in employing automated methods of inference to bioactive molecule discovery is remarkable, the fields of machine learning and deep learning are rapidly advancing as well. However, many machine learning methods for bioactive ligand discovery still rely on traditional fingerprint representations [55,60] or other molecule descriptors that are being derived manually [54]. Currently efforts are also being made to move beyond the analysis of static structures and combine molecular dynamics with machine learning for predicting bioactive GPCR ligands and distinguishing between antagonists and agonists [87]. However, whether these methods can further be improved by considering three-dimensional, non-static representations of molecules remains to be explored.

## Declaration of interest

None.

## Acknowledgements

This project was supported by the University of Wisconsin-Madison College of Letters & Science. The author thanks Drs. Leslie A. Kuhn, Spencer S. Ericksen, and Anthony Gitter for helpful feedback on the manuscript.

# References and recommended reading

• of special interest
•• of outstanding interest

•1. Hauser AS, Attwood MM, Rask-Andersen M, Schiöth HB, Gloriam DE: **Trends in GPCR drug discovery: new agents, targets and indications**. *Nat Rev Drug Discov* 2017, **16**:829–842.

2. Garland SL: **Are GPCRs still a source of new targets?** *J Biomol Screen* 2013, **18**:947–966.

3. Thomsen W, Frazer J, Unett D: **Functional assays for screening GPCR targets**. *Curr Opin Biotech* 2005, **16**:655–665.

4. Bjarnadóttir TK, Gloriam DE, Hellstrand SH, Kristiansson H, Fredriksson R, Schiöth HB: **Comprehensive repertoire and phylogenetic analysis of the G protein-coupled receptors in human and mouse**. *Genomics* 2006, **88**:263–273.

5. Davenport AP, Alexander SPH, Sharman JL, Pawson AJ, Benson HE, Monaghan AE, Liew WC, Mpamhanga CP, Bonner TI, Neubig RR, et al.: **International Union of Basic and Clinical Pharmacology. LXXXVIII. G protein-coupled receptor list: recommendations for new pairings with cognate ligands**. *Pharmacol Rev* 2013, **65**:967–986.

6. Southan C, Sharman JL, Benson HE, Faccenda E, Pawson AJ, Alexander SPH, Buneman OP, Davenport AP, McGrath JC, Peters JA, et al.: **The IUPHAR/BPS guide to PHARMACOLOGY in 2016: towards curated quantitative interactions between 1300 protein targets and 6000 ligands**. *Nucleic Acids Res* 2016, **44**:D1054–D1068.

7. Sterling T, Irwin JJ: **ZINC 15–ligand discovery for everyone**. *J Chem Inf Model* 2015, **55**:2324–2337.

8. Sunseri J, Koes DR: **Pharmit: interactive exploration of chemical space**. *Nucleic Acids Res* 2016, **44**:W442–W448.

9. Bento AP, Gaulton A, Hersey A, Bellis LJ, Chambers J, Davies M, Krüger FA, Light Y, Mak L, McGlinchey S, et al.: **The ChEMBL bioactivity database: an update**. *Nucleic Acids Res* 2014, **42**:D1083–D1090.

10. Lipinski CA: **Drug-like properties and the causes of poor solubility and poor permeability**. *J Pharmacol Tox Met* 2000, **44**:235–249.

11. Baell J, Walters MA: **Chemistry: chemical con artists foil drug discovery**. *Nature News* 2014, **513**:481.

12. Van De Waterbeemd H, Gifford E: **ADMET in silico modelling: towards prediction paradise?** *Nat Rev Drug Discov* 2003, **2**:192–204.

13. Liu W, Wacker D, Wang C, Abola E, Cherezov V: **Femtosecond crystallography of membrane proteins in the lipidic cubic phase**. *Philos T Roy Soc B* 2014, **369**:20130314–20130314.

14. Renaud J-P, Chari A, Ciferri C, Liu W, Rémigy H-W, Stark H, Wiesmann C: **Cryo-EM in drug discovery: achievements, limitations and prospects**. *Nat Rev Drug Discov* 2018, **17**:471–492.

15. Basith S, Cui M, Macalino SJY, Park J, Clavio NAB, Kang S, Choi S: **Exploring G protein-coupled receptors (GPCRs) ligand space via cheminformatics approaches: impact on rational drug design**. *Front Pharmacol* 2018, **9**:1–26.

16. Wheatley M, Wootten D, Conner M, Simms J, Kendrick R, Logan R, Poyner D, Barwell J: **Lifting the lid on GPCRs: the role of extracellular loops: GPCR extracellular loops**. *Brit J Pharmacol* 2012, **165**:1688–1703.

•17. Raschka S, Scott AM, Liu N, Gunturu S, Huertas M, Li W, Kuhn LA: **Enabling the hypothesis-driven prioritization of ligand candidates in big databases: Screenlamp and its application to GPCR inhibitor discovery for invasive species control**. *J Comput Aid Mol Des* 2018, **32**:415–433.

18. Raschka S, Mirjalili V: *Python Machine Learning, 2nd Ed.* Packt Publishing; 2017.

••19. Raschka S, Scott AM, Huertas M, Li W, Kuhn LA: **Automated inference of chemical discriminants of biological activity**. In *Computational Drug Discovery and Design*. Springer New York; 2018:307–338.

20. Liang Y-L, Zhao P, Draper-Joyce C, Baltos J-A, Glukhova A, Truong TT, May LT, Christopoulos A, Wootten D, Sexton PM, et al.: **Dominant negative G proteins enhance formation and purification of agonist-GPCR-G protein complexes for structure determination**. *ACS Pharmacol Transl* 2018, **1**:12–20.

21. Klabunde T, Hessler G: **Drug design strategies for targeting G-protein-coupled receptors**. *ChemBioChem* 2002, **3**:928–944.

22. Flohr S, Kurz M, Kostenis E, Brkovich A, Fournier A, Klabunde T: **Identification of nonpeptidic urotensin II receptor antagonists by virtual screening based on a pharmacophore model derived from structure- activity relationships and nuclear magnetic resonance studies on urotensin II**. *J Med Chem* 2002, **45**:1799–1805.

23. Evers A, Hessler G, Matter H, Klabunde T: **Virtual screening of biogenic amine-binding G-protein coupled receptors: comparative evaluation of protein-and ligand-based virtual screening protocols**. *J Med Chem* 2005, **48**:5448–5465.

24. Bender A, Glen RC: **Molecular similarity: a key technique in molecular informatics**. *Org Biomol Chem* 2004, **2**:3204–3218.

25. Bender A, Jenkins JL, Scheiber J, Sukuru SCK, Glick M, Davies JW: **How similar are similarity searching methods? A principal component analysis of molecular descriptor space**. *J Chem Inf Model* 2009, **49**:108–119.

26. Bajusz D, Rácz A, Héberger K: **Why is Tanimoto index an appropriate choice for fingerprint-based similarity calculations?** *J Cheminformatics* 2015, **7**:20.

27. Duan J, Dixon SL, Lowrie JF, Sherman W: **Analysis and comparison of 2D fingerprints: insights into database screening performance using eight fingerprint methods**. *J Mol Graph Model* 2010, **29**:157–170.

28. Willett P: **Similarity-based virtual screening using 2D fingerprints**. *Drug Discov Today* 2006, **11**:1046–1053.

29. Awale M, Jin X, Reymond J-L: **Stereoselective virtual screening of the ZINC database using atom pair 3D-fingerprints**. *J Cheminformatics* 2015, **7**:3.

30. Hawkins PCD, Skillman AG, Nicholls A: **Comparison of shape-matching and docking as virtual screening tools**. *J Med Chem* 2007, **50**:74–82.

•31. Cortés-Ciriano I, Firth NC, Bender A, Watson O: **Discovering highly potent molecules from an initial set of inactives using iterative screening**. *J Chem Inf Model* 2018, **58**:2000–2014.

32. Ngo T, Kufareva I, Coleman JL, Graham RM, Abagyan R, Smith NJ: **Identifying ligands at orphan GPCRs: current status using structure-based approaches: approaches for identifying orphan GPCR ligands**. *Brit J Pharmacol* 2016, **173**:2934–2951.

•33. Ciancetta A, Cuzzolin A, Deganutti G, Sturlese M, Salmaso V, Cristiani A, Sabbadin D, Moro S: **New trends in inspecting GPCR-ligand recognition process: the contribution of the molecular modeling section (MMS) at the University of Padova**. *Mol Inform* 2016, **35**:440–448.

34. Sengupta D, Sonar K, Joshi M: **Characterizing clinically relevant natural variants of GPCRs using computational approaches**. In *Methods in Cell Biology*. Elsevier; 2017:187–204.

35. Hadianawala M, Mahapatra AD, Yadav JK, Datta B: **Molecular docking, molecular modeling, and molecular dynamics studies of azaisoflavone as dual COX-2 inhibitors and TP receptor antagonists**. *J Mol Model* 2018, **24**:69.

36. Dacanay F, Ladra M, Junio H, Nellas R: **Molecular affinity of mabolo extracts to an octopamine receptor of a fruit fly**. *Molecules* 2017, **22**:1677.

37. Raschka S, Bemister-Buffington J, Kuhn LA: **Detecting the native ligand orientation by interfacial rigidity: SiteInterlock**. *Proteins* 2016, **84**:1888–1901.

•38. Schneider J, Korshunova K, Musiani F, Alfonso-Prieto M, Giorgetti A, Carloni P: **Predicting ligand binding poses for low-resolution membrane protein models: perspectives from multiscale simulations**. *Biochem Bioph Res Co* 2018, **498**:366–374.

39. Roth BL, Irwin JJ, Shoichet BK: **Discovery of new GPCR ligands to illuminate new biology**. *Nat Chem Bio* 2017, **13**:1143–1151.

40. Jakubík J, El-Fakahany EE, Doležal V: **Towards predictive docking at aminergic G-protein coupled receptors**. *J Mol Model* 2015, **21**:284.

41. Chen H, Engkvist O, Wang Y, Olivecrona M, Blaschke T: **The rise of deep learning in drug discovery**. *Drug Discov Today* 2018, **23**:1241–1250.

42. Ekins S, Clark AM, Swamidass SJ, Litterman N, Williams AJ: **Bigger data, collaborative tools and the future of predictive drug discovery**. *J Comput Aid Mol Des* 2014, **28**:997–1008.

43. Lavecchia A: **Machine-learning approaches in drug discovery: methods and applications**. *Drug Discov Today* 2015, **20**:318–331.

44. Lima AN, Philot EA, Trossini GHG, Scott LPB, Maltarollo VG, Honorio KM: **Use of machine learning approaches for novel drug discovery**. *Expert Opin Drug Dis* 2016, **11**:225–239.

45. Wale N: **Machine learning in drug discovery and development**. *Drug Develop Res* 2011, **72**:112–119.

46. Ekins S: **The next era: deep learning in pharmaceutical research**. *Pharm Res* 2016, **33**:2594–2603.

•47. Ching T, Himmelstein DS, Beaulieu-Jones BK, Kalinin AA, Do BT, Way GP, Ferrero E, Agapow P-M, Zietz M, Hoffman MM, et al.: **Opportunities and obstacles for deep learning in biology and medicine**. *J Roy Soc Interface* 2018, **15**:20170387.

•48. Ericksen SS, Wu H, Zhang H, Michael LA, Newton MA, Hoffmann FM, Wildman SA: **Machine learning consensus scoring improves performance across targets in structure-based virtual screening**. *J Chem Inf Model* 2017, **57**:1579–1590.

49. Wang C, Liu J, Luo F, Tan Y, Deng Z, Hu Q-N: **Pairwise input neural network for target-ligand interaction prediction**. In *2014 IEEE International Conference on Bioinformatics and Biomedicine (BIBM)*. IEEE; 2014:67–70.

50. Raschka S: **MLxtend: Providing machine learning and data science utilities and extensions to Python's scientific computing stack**. *J Open Source Softw* 2018, **3**:638.

51. Ribeiro MT, Singh S, Guestrin C: **"Why should I trust you?": explaining the predictions of any classifier**. In *Proceedings of the 22nd ACM SIGKDD International Conference on Knowledge Discovery and Data Mining - KDD '16*. ACM Press; 2016:1135–1144.

52. Jacobs DJ, Rader AJ, Kuhn LA, Thorpe MF: **Protein flexibility predictions using graph theory**. *Proteins* 2001, **44**:150–165.

•53. Ghasemi F, Mehridehnavi A, Pérez-Garrido A, Pérez-Sánchez H: **Neural network and deep-learning algorithms used in QSAR studies: merits and drawbacks**. *Drug Discov Today* 2018, **23**:1784–1790.

54. Ma J, Sheridan RP, Liaw A, Dahl GE, Svetnik V: **Deep neural nets as a method for quantitative structure–activity relationships**. *J Chem Inf Model* 2015, **55**:263–274.

•55. Koutsoukas A, Monaghan KJ, Li X, Huan J: **Deep-learning: investigating deep neural networks hyper-parameters and comparison of performance to shallow methods for modeling bioactivity data**. *J Cheminformatics* 2017, **9**:42.

56. Lewis DD: **Naive (Bayes) at forty: the independence assumption in information retrieval**. In *Machine Learning: ECML-98*. Edited by Nédellec C, Rouveirol C. Springer Berlin Heidelberg; 1998:4–15.

57. Wang S, Sun H, Liu H, Li D, Li Y, Hou T: **ADMET evaluation in drug discovery. 16. Predicting hERG blockers by combining multiple pharmacophores and machine learning approaches**. *Mol Pharm* 2016, **13**:2855–2866.

58. Nidhi, Glick M, Davies JW, Jenkins JL: **Prediction of biological targets for compounds using multiple-category Bayesian models trained on chemogenomics databases**. *J Chem Inf Model* 2006, **46**:1124–1133.

59. Bender A, Scheiber J, Glick M, Davies JW, Azzaoui K, Hamon J, Urban L, Whitebread S, Jenkins JL: **Analysis of pharmacology data and the prediction of adverse drug reactions and off-target effects from chemical structure**. *Chem Med Chem* 2007, **2**:861–873.

60. Clark AM, Ekins S: **Open source Bayesian models. 2. Mining a "big dataset" to create and validate models with ChEMBL**. *J Chem Inf Model* 2015, **55**:1246–1260.

61. Hosmer Jr DW, Lemeshow S, Sturdivant RX: *Applied Logistic Regression*. John Wiley & Sons; 2013.

62. Caster O, Norén GN, Madigan D, Bate A: **Large-scale regression-based pattern discovery: the example of screening the WHO global drug safety database**. *Statistical Analy Data Mining* 2010, **3**:197–208.

63. Harpaz R, DuMouchel W, Shah NH, Madigan D, Ryan P, Friedman C: **Novel data-mining methodologies for adverse drug event discovery and analysis**. *J Clin Pharm Ther* 2012, **91**:1010–1021.

64. Kim S, Jin D, Lee H: **Predicting drug-target interactions using drug-drug interactions**. *PLoS ONE* 2013, **8**:e80129.

65. Cristianini N, Shawe-Taylor J, others: *An introduction to support vector machines and other kernel-based learning methods*. Cambridge University Press; 2000.

66. Kriegl JM, Arnhold T, Beck B, Fox T: **A support vector machine approach to classify human cytochrome P450 3A4 inhibitors**. *J Comput Aid Mol Des* 2005, **19**:189–201.

67. Breiman L: **Random forests**. *Mach Learn* 2001, **45**:5–32.

68. Svetnik V, Liaw A, Tong C, Culberson JC, Sheridan RP, Feuston BP: **Random forest: a classification and regression tool for compound classification and QSAR modeling**. *J Chem Inf Model* 2003, **43**:1947–1958.

69. Svetnik V, Liaw A, Tong C, Wang T: **Application of Breiman's random forest to modeling structure-activity relationships of pharmaceutical molecules**. *International Workshop on Multiple Classifier Systems*. 2004:334–343.

70. Rakers C, Reker D, Brown JB: **Small random forest models for effective chemogenomic active learning**. *J Med Chem* 2017, **18**:124–142.

•71. Korotcov A, Tkachenko V, Russo DP, Ekins S: **Comparison of deep learning with multiple machine learning methods and metrics using diverse drug discovery data sets**. *Mol Pharm* 2017, **14**:4462–4475.

72. Lenselink EB, ten Dijke N, Bongers B, Papadatos G, van Vlijmen HWT, Kowalczyk W, IJzerman AP, van Westen GJP: **Beyond the hype: deep neural networks outperform established methods using a ChEMBL bioactivity benchmark set**. *J Cheminformatics* 2017, **9**:45.

•73. Mayr A, Klambauer G, Unterthiner T, Steijaert M, Wegner JK, Ceulemans H, Clevert D-A, Hochreiter S: **Large-scale comparison of machine learning methods for drug target prediction on ChEMBL**. *Chem Sci* 2018, **9**:5441–5451.

•74. Russo DP, Zorn KM, Clark AM, Zhu H, Ekins S: **Comparing multiple machine learning algorithms and metrics for estrogen receptor binding prediction**. *Mol Pharm* 2018, **15**:4361–4370.

•75. Liu S, Alnammi M, Ericksen SS, Voter AF, Ananiev GE, Keck JL, Hoffmann FM, Wildman SA, Gitter A: **Practical model selection for prospective virtual screening**. *J Chem Inf Model* 2018,

76. Duvenaud D, Maclaurin D, Aguilera-Iparraguirre J, Gómez-Bombarelli R, Hirzel T, Aspuru-Guzik A, Adams RP: **Convolutional networks on graphs for learning molecular fingerprints**. In *Proceedings of the 28th International Conference on Neural Information Processing Systems*. 2015.

••77. Hop P, Allgood B, Yu J: **Geometric deep learning autonomously learns chemical features that outperform those engineered by domain experts**. *Mol Pharm* 2018, **15**:4371–4377.

•78. Zhavoronkov A: **Artificial intelligence for drug discovery, biomarker development, and generation of novel chemistry**. *Mol Pharm* 2018, **15**:4311–4313.

79. Raschka S: **BioPandas: working with molecular structures in pandas DataFrames**. *J Open Source Softw* 2017, **2**:279.

80. Ochoa R, Davies M, Papadatos G, Atkinson F, Overington JP: **myChEMBL: a virtual machine implementation of open data and cheminformatics tools**. *Bioinformatics* 2013, **30**:298–300.

81. Cock PJ, Antao T, Chang JT, Chapman BA, Cox CJ, Dalke A, Friedberg I, Hamelryck T, Kauff F, Wilczynski B, et al.: **Biopython: freely available Python tools for computational molecular biology and bioinformatics**. *Bioinformatics* 2009, **25**:1422–1423.

82. Pedregosa F, Varoquaux G, Gramfort A, Michel V, Thirion B, Grisel O, Blondel M, Prettenhofer P, Weiss R, Dubourg V, et al.: **Scikit-learn: Machine learning in Python**. *J Mach Learn Res* 2011, **12**:2825–2830.

83. Abadi M, Barham P, Chen J, Chen Z, Davis A, Dean J, Devin M, Ghemawat S, Irving G, Isard M, et al.: **TensorFlow: a system for large-scale machine learning**. In *OSDI*. 2016:265–283.

84. Paszke A, Gross S, Chintala S, Chanan G, Yang E, DeVito Z, Lin Z, Desmaison A, Antiga L, Lerer A: **Automatic differentiation in PyTorch**. In *Proceedings of the 30th International Conference on Neural Information Processing Systems*. 2017:4.

•85. Chuang KV, Keiser MJ: **Adversarial controls for scientific machine learning**. *ACS Chem Biol* 2018, **13**:2819–2821.

86. Raschka S, Wolf AJ, Bemister-Buffington J, Kuhn LA: **Protein–ligand interfaces are polarized: discovery of a strong trend for intermolecular hydrogen bonds to favor donors on the protein side with implications for predicting and designing ligand complexes**. *J Comput Aid Mol Des* 2018, **32**:511–528.

87. Feinberg EN, Farimani AB, Uprety R, Hunkele A, Pasternak GW, Majumdar S, Pande VS: **Machine Learning Harnesses Molecular Dynamics to Discover New μ Opioid Chemotypes**. *arXiv preprint arXiv:180304479* 2018,